\documentclass[preprint,number,5p,times,twocolumn]{elsarticle}

\usepackage{amssymb}
\usepackage{lipsum}
\usepackage{subcaption} 
\usepackage{stfloats}  
\usepackage{float} 
\usepackage{graphicx}
\usepackage{overpic} 
\usepackage{textcomp}
\usepackage{gensymb}
\usepackage{hyperref}
\usepackage{url}
\usepackage{lipsum}
\usepackage{amsmath}
\usepackage{amsfonts}

\usepackage{setspace}
\usepackage{hyperref}

\usepackage{bm}
\usepackage{svg}
\usepackage{array}
\usepackage{graphicx}
\usepackage{textcomp}

\usepackage{xcolor}

\hypersetup{
colorlinks=true,
linkcolor=blue,
filecolor=magenta,
urlcolor=blue
}

\journal{arXiv}

\begin{document}

\begin{frontmatter}



\title{Electron doping in single crystalline BaBiO$_3$: BaBiO$_{3-x}$F$_x$}


\author[inst1]{Sathishkumar M\fnref{fn1}}
\author[inst1]{Asha Ann Abraham}
\author[inst2,inst3]{Rajesh Kumar Sahu}
\author[inst2,inst3]{Soma Banik}
\author[inst1]{Soham Manni\corref{cor1}}
\address[inst1]{Department of Physics, Indian Institute of Technology Palakkad, Kerala - 678623, India}
\address[inst2]{Accelerator Physics and Synchrotrons Utilization Division, Raja Ramanna Centre for Advanced Technology, Indore 452013, India}
\address[inst3]{Homi Bhabha National Institute, Training School Complex, Anushakti Nagar, Mumbai 400094, India}

\cortext[cor1]{\href{mailto:smanni@iitpkd.ac.in}{smanni@iitpkd.ac.in}}
\fntext[fn1]{\href{mailto:222304002@smail.iitpkd.ac.in}{222304002@smail.iitpkd.ac.in}}
\begin{abstract}
Topological insulators are a new class of insulators with conducting surface state. Most of the topological insulators are chalcogenides, where a tiny amount of chalcogen vacancy destroys the predicted bulk insulating state and results in a metallic or semi-metallic bulk electrical transport.   BaBiO$_3$ (BBO) is an interesting large bandgap (0.7 eV) insulator that upon hole doping becomes a superconductor and is theoretically predicted to show a topological insulating state under electron doping. We have explored electron doping through the chemical substitution of fluorine atoms at the oxygen site. The single crystals of BBO and fluorine-doped BBO were synthesized via a one-step solid-state technique. The single crystals of pure BBO and 10 $\%$ F -doped BBO (BaBiO$_{2.7}$F$_{0.3}$) are chemically single-phase samples and crystallize in monoclinic \textit{I2/m} crystal structure.  The core level and valence band X-ray photoelectron spectra confirm electron doping in the 10$\%$ fluorine-doped BBO. 20 $\%$ F-doped BBO appears to be a multiphase sample, confirmed by back-scattered electron (BSE) imaging and X-ray diffraction. This article reports on the successful growth of pure and F-doped BBO using a one-step solid-state technique and discusses the effect of F-doping on structural and electronic properties.

\end{abstract}

\end{frontmatter}

\section{Introduction}
\label{introduction}
In condensed matter physics, materials have been classified as metals, semiconductors, and insulators based on their transport properties. In metals, the conduction and valence bands overlap, whereas in insulators, the conduction and valence bands are gapped. Topological materials (TMs) are a class of materials that have a topologically non-trivial electronic band structure, e.g., conducting surface state, Dirac point, Weyl point, Fermi arc, etc. Topology in mathematics deals with properties that are invariant under smooth or continuous deformations.  The topological insulator (TI), a special class of TM that has a bulk insulating state and a conducting surface state and is protected against mild perturbations. Kane and Mele predicted the topological insulating state in graphene, but the spin-orbit coupling in graphene is not large enough to realize the topological insulating state in graphene experimentally\cite{kane2005quantum}. The surface states in TI disperse linearly in momentum space and are spin-momentum locked. This facilitates dissipationless channels for transport, as the electrons cannot backscatter. The topological insulating state was first experimentally observed in HgTe quantum well sandwiched between CdTe layers by Koning \textit{ et al.} \cite{konig2007quantum}. This marks the first two-dimensional topological insulator. The idea was extended to three dimensions. The first three-dimensional (3D) topological insulator observed experimentally is the Bi$_{1-x}$Sb$_x$\cite{hasan2011three}. There are many 3D bismuth chalcogenide topological insulators, including Bi$_2$Te$_3$ and Bi$_2$Se$_3$. The Bi$_2$Te$_3$ in its pristine form without vacancies or defects is a pure topological insulator with a bulk band gap. However, these materials are difficult to grow in pristine form without defects with a stoichiometric composition. The chalcogen vacancies in such materials cause unintentional doping, so the Fermi level lies above the bottom of the bulk conduction band and does not cross the gapless surface states. They show semimetallic behavior in resistivity measurement, making it difficult to achieve charge conduction only from the surface states. A topological insulator with a large bulk band gap ensures that the conduction originates only from the surface states and not from the impurity states caused by unintentional doping or vacancies. The largest bulk band gap in the known topological insulators is around 0.3 eV. These shortcomings have interested researches to look for topological insulators with a large bulk band gap.

Yan \textit{et al.}\cite{yan2013large} proposed that electron-doped BaBiO$_3$ can be a topological insulator under a large electron doping level of one electron per unit cell. Initially, pure BaBiO$_3$ (BBO) was predicted to show metallic behavior due to half-filled Bismuth 6s-bands \cite{sleight2015bismuthates}. However, experimentally insulating behavior was observed with a large optical band gap of 2.25 eV \cite{chouhan2018babio3}. The reason for the insulating behavior was studied for a long time. The insulating behavior was proposed to arise  from the breathing mode of oxygen due to the Bi charge-disproportionation into $Bi^{3+}$ and $Bi^{5+}$ causing a bond-disproportionation \cite{bouwmeester2021babio3}.The breathing mode can also arise from the hybridization of the Bi-6s and O-2p orbitals, resulting in the insulating behavior.  BaBiO$_3$, when doped with holes, becomes a superconductor. The hole doping in BBO with potassium \cite{cava1988superconductivity} and lead \cite{sleight1993high} resulted in a superconducting transition temperature higher than that reported in some of the cuprates and arises due to the suppression of the breathing mode.

BBO, having a large indirect band gap of 0.7 eV \cite{yan2013large} in its topological insulating state, is an ideal candidate among the predicted topological insulators, as conduction can arise purely from the surface states due to its large bulk band gap. The topological insulating state in BBO can be realized by shifting the Fermi level by 2eV so that the surface states lie at the Fermi level. The Fermi level can be shifted up by electron doping. The electron doping mechanism in BBO has not been widely explored. Simulation studies carried out by Khamari\textit{ et al.}\cite{khamari2019shifting} predict that the topological insulating phase in BBO can be achieved by substituting 33$\%$ of oxygen atoms with fluorine atoms. Despite this promising potential, fluorine doping via solid-state methods has proven challenging because of the high volatility of fluorine. There have been no reports on single-crystal growth of fluorine-doped BBO via solid-state technique in the literature. Recently, Vesto \textit{ et al.} \cite{vesto2025growth} reported electron doping in fluorine-doped BBO thin films on polished sapphire substrates.
 
 We have successfully synthesized bulk fluorine-doped BBO (F-doped BBO). The crystals were chemically and structurally characterized. Their core and valence band electronic spectra were studied to confirm electron doping. The limit of fluorine doping in the bulk BBO crystal was also determined. 

\section{Experimental details}
\begin{figure}[H]
  \centering
  \begin{subfigure}[b]{0.31\linewidth}
    \begin{overpic}[width=\linewidth]{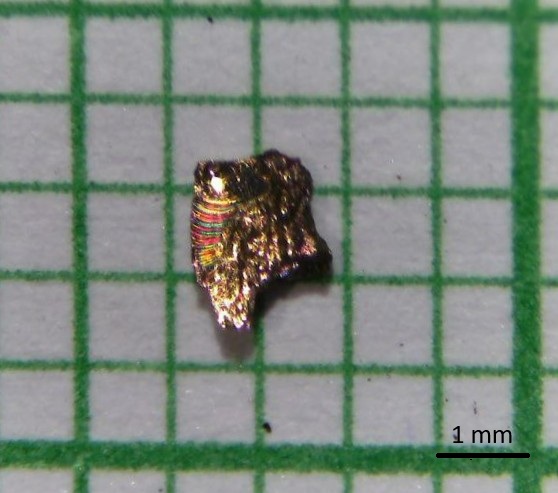}
     \put(3,75){\textcolor{white}{(a)}}
    \end{overpic}
\phantomcaption
\label{fig:1a}
  \end{subfigure}
  \hfill
  \begin{subfigure}[b]{0.3\linewidth}
    \begin{overpic}[width=\linewidth]{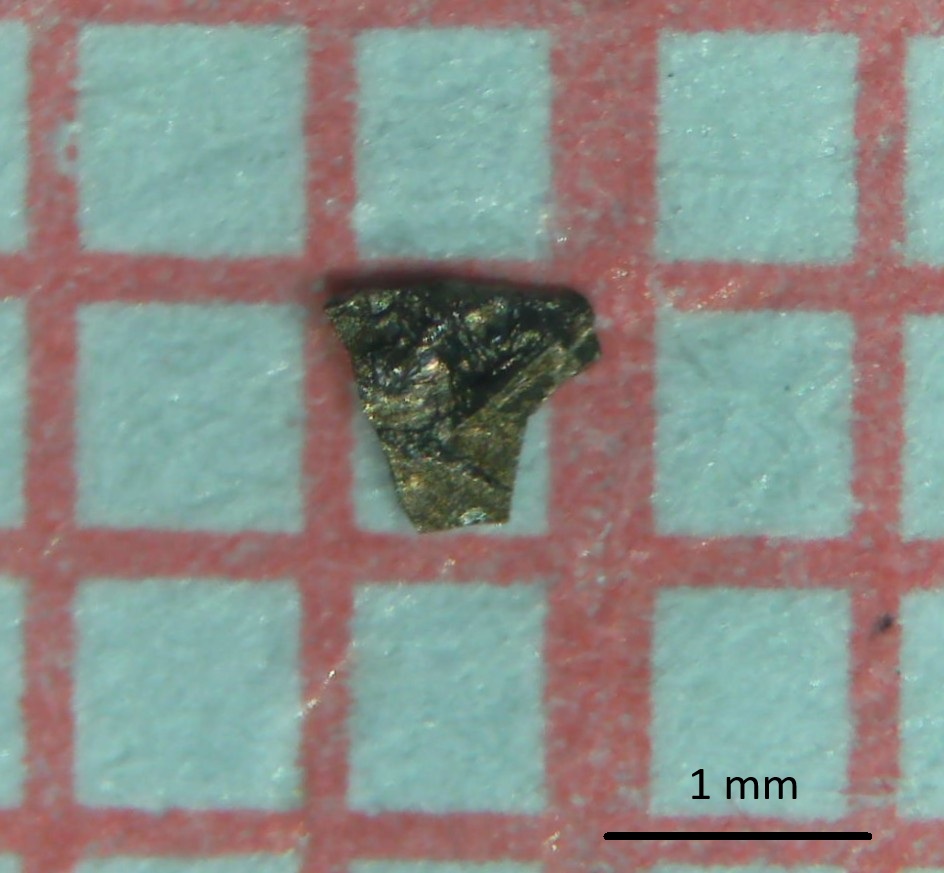}
     \put(3,75){\textcolor{white}{(b)}}
    \end{overpic}
\phantomcaption
\label{fig:1b}
  \end{subfigure}
  \hfill
  \begin{subfigure}[b]{0.28\linewidth}
    \begin{overpic}[width=\linewidth]{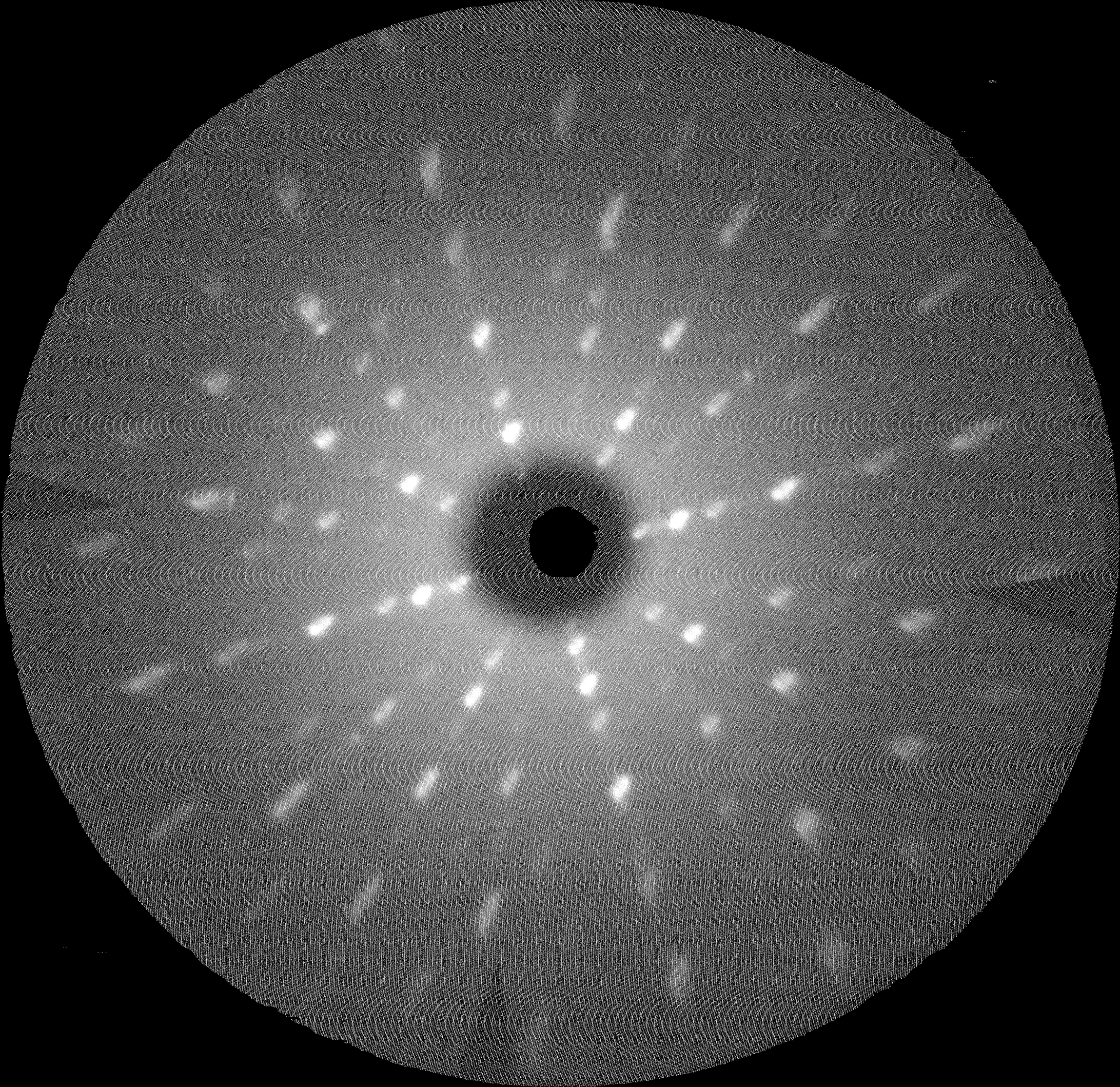}
     \put(3,80){\textcolor{white}{(c)}}
    \end{overpic}
\phantomcaption
\label{fig:1c}
  \end{subfigure}
  \caption{(a, b) Photographs of pure BBO and 10 $\%$ F-doped BBO single crystals, respectively, and (c) Laue diffraction pattern of the pure BBO single crystals along [001] direction  }
\end{figure}

BaBiO$_3$ is a congruently melting material. Earlier single crystals of this material and its hole-doped counterparts were grown by slow cooling the melt of the stoichiometric polycrystal. Polycrystals were synthesized mainly by a solid-state reaction of stoichiometric carbonates and Bi$_2$O$_3$ \cite{shiozaki1993crystal}. We have successfully grown pure BBO crystals following this technique. Since our goal was to synthesize F-doped BBO in single-crystalline form, we wanted to avoid the polycrystalline step, which often results in secondary impurity phases at the time of doping.  The single crystals of BBO and F-doped BBO were synthesized using a one-step solid-state technique. The precursors were  BaO (99.99$\%$ metal basis, Alfa Aesar), Bi$_2$O$_3$ (99.975$\%$ metal basis, Alfa Aesar) for undoped BBO and, for F-doped BBO, additionally BiF$_3$ (99.99$\%$ metal basis, Alfa Aesar) was used. We have attempted to grow undoped BBO (BaBiO$_3$), 10$\%$ Fluorine doped BBO (BaBiO$_{2.4}$F$_{0.6}$) and 20$\%$ Fluorine doped BBO (BaBiO$_{2.7}$F$_{0.3}$). The stoichiometric mixture of precursors was ground in a mortar and pestle, and the mixture was placed in an alumina crucible. The ground precursors were heated in open air to 1100 \degree C and held for 24 hours to form a homogeneous melt and slowly cooled to 1015 \degree C at a rate of 1 \degree C / hour and then to room temperature by furnace cooling. Single crystals were extracted by breaking the crucible and mechanically separating pieces from the solidified melt for electron microscopy and photoelectron spectroscopy. Figure [\ref{fig:1a},\ref{fig:1b}] show pictures of pure and F-doped BBO crystals. Because of the perovskite structure and congruently melting compound, it is hard to cleave a crystal with a clear single-crystalline facet.

The grown crystals were structurally characterized via X-ray diffraction (XRD) on the powdered single crystals extracted from the crucible. XRD was performed using a Rigaku SmartLab 9 kW system with Cu K$_{\alpha}$ ($\lambda$ = 1.54 $\AA$) radiation at room temperature from 10 to 100 degree at a step size of 0.01 $\degree$ in Bragg-Brentano geometry. The XRD pattern was analyzed using the Rietveld refinement technique with General Structure and Analysis software (GSAS-$\mathrm{II}$) to determine the crystal structure and lattice parameters \cite{Toby:aj5212}. The Back scattered electron (BSE) images were taken using a Carl Zeiss Gemini SEM 300 field emission scanning electron microscope (FESEM).

Synchrotron X-ray photoelectron spectroscopy (XPS) measurements were performed at the angle-resolved photoelectron spectroscopy beamline (BL-10), Indus 2 at RRCAT, Indore, on the single crystals of BBO and 10\% F-doped BBO crystals. The core level and valence band spectra were measured using a photon energy (h$\nu$) of 838.4 eV, with an energy resolution of 0.37 eV, using the SPECS Phoibos 150 electron analyser. The crystals were sputtered with Ar ions with an energy of 1.5 keV for 5 minutes before the experiment to ensure a clean surface. The base vacuum during the photoemission measurements was 7 $\times$ 10$^{-11}$ mbar.

\section{Results and Discussion}
\begin{figure}[H]
    \centering
    \includegraphics[width=\linewidth]{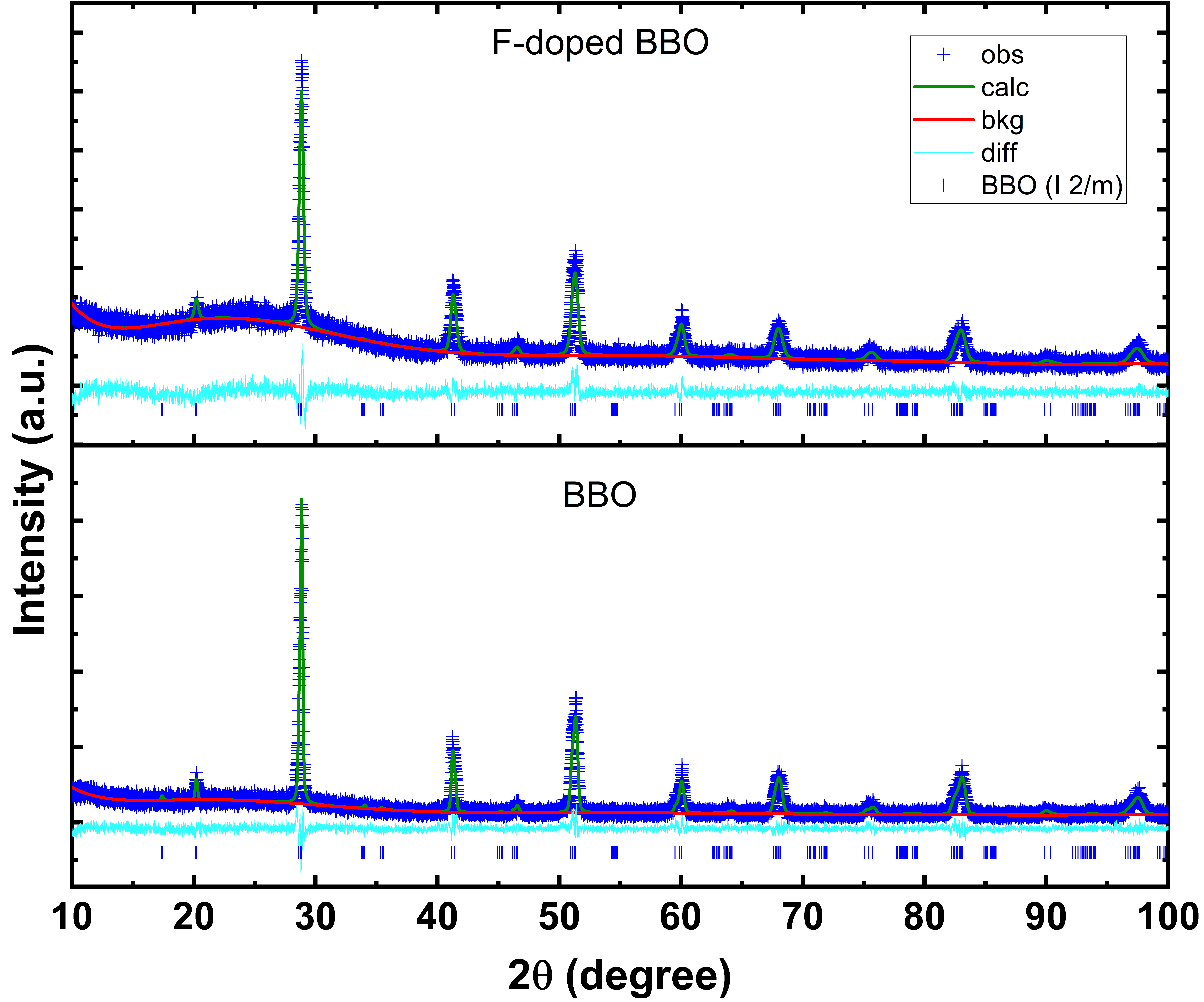}
    \caption{Rietveld refinement of undoped BBO (bottom panel) and 10\% F-doped BBO (top panel) where, $obs$ is  experimental data from XRD , $calc$ is simulated data from Rietveld refinement, $bkg$ is background fit, $diff$ is the difference between the calculated and measured data and the vertical blue lines are BaBiO$_3$ - Bragg peak positions.}
    \label{Refinement}
\end{figure}

Figure [\ref{fig:1c}] shows the Laue diffraction of the pure BBO crystal, which confirms the single crystallinity of the sample. Cleaved F-doped crystals were smaller than 1mm x 1mm, so it was difficult to get a good Laue image from a single-crystalline piece. Powder X-ray diffraction (PXRD) was performed on the powdered crystals that were mechanically separated from the melt. A large chunk of the solidified melt was also powdered for PXRD measurement to confirm the single phase of the full growth in the crucible. The PXRD data for pure and 10$\%$ F-doped BBO were refined with a monoclinic $I2 / m$ crystal structure (SG No. 12), shown in Figure [\ref{Refinement}]. It can be confirmed that both pure BBO and 10$\%$ F-doped BBO are single phase because there are no impurity peaks present.  The refined lattice parameters obtained from the Rietveld refinement are listed in Table [\ref{tab:1}] and the values for pure BBO are comparable to the reported values \cite{pei1990structural}. Only a very slight change in the lattice parameters is observed from lab XRD due to 10$\%$ fluorine doping. It is hard to refine the fluorine occupancy from the refinement; hence, the nominal composition was used to populate the oxygen sites. The refined atomic coordinates for pure and F-doped BBO are listed in Table [\ref{AtomicCoordinates}]. The PXRD pattern of 20$\%$ F-doped BBO sample shown in Figure [\ref{fig:A5}], clearly indicates the presence of an impurity phase along with the BBO phase in the sample.

\begin{figure*}[b]
  \centering
  \begin{subfigure}[b]{0.33\textwidth}
    \begin{overpic}[width=\linewidth]{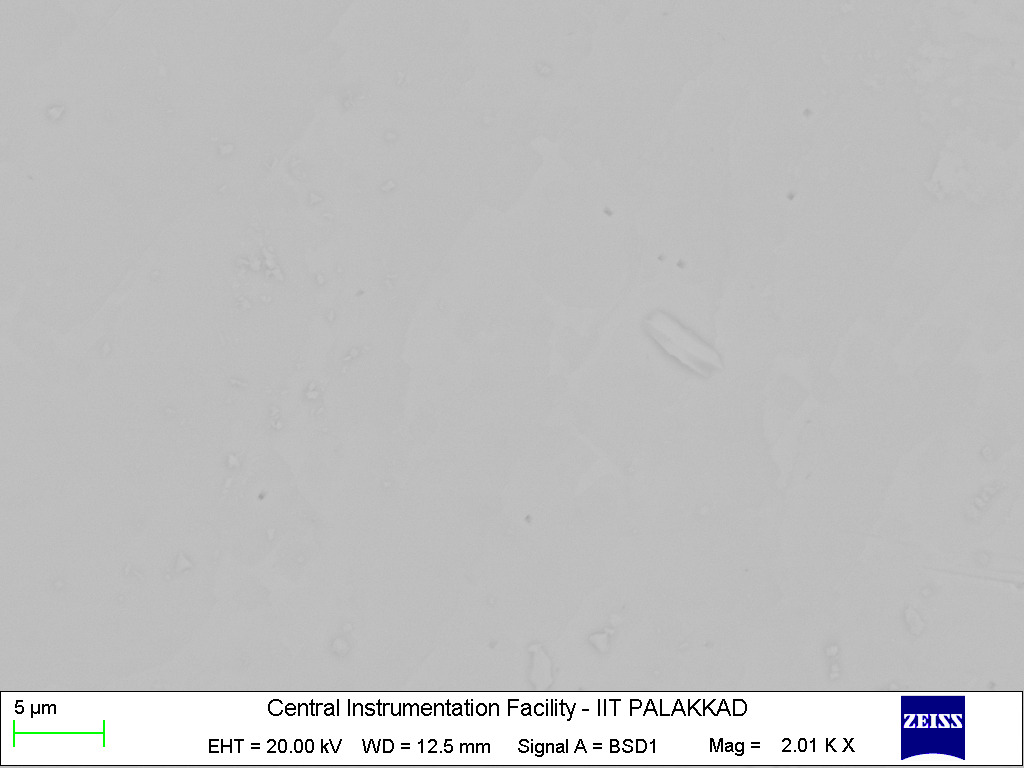}
     \put(3,68){\textcolor{black}{(a)}}
    \end{overpic}
\phantomcaption
\label{fig:3a}
  \end{subfigure}
  \hfill
  \begin{subfigure}[b]{0.33\textwidth}
    \begin{overpic}[width=\linewidth]{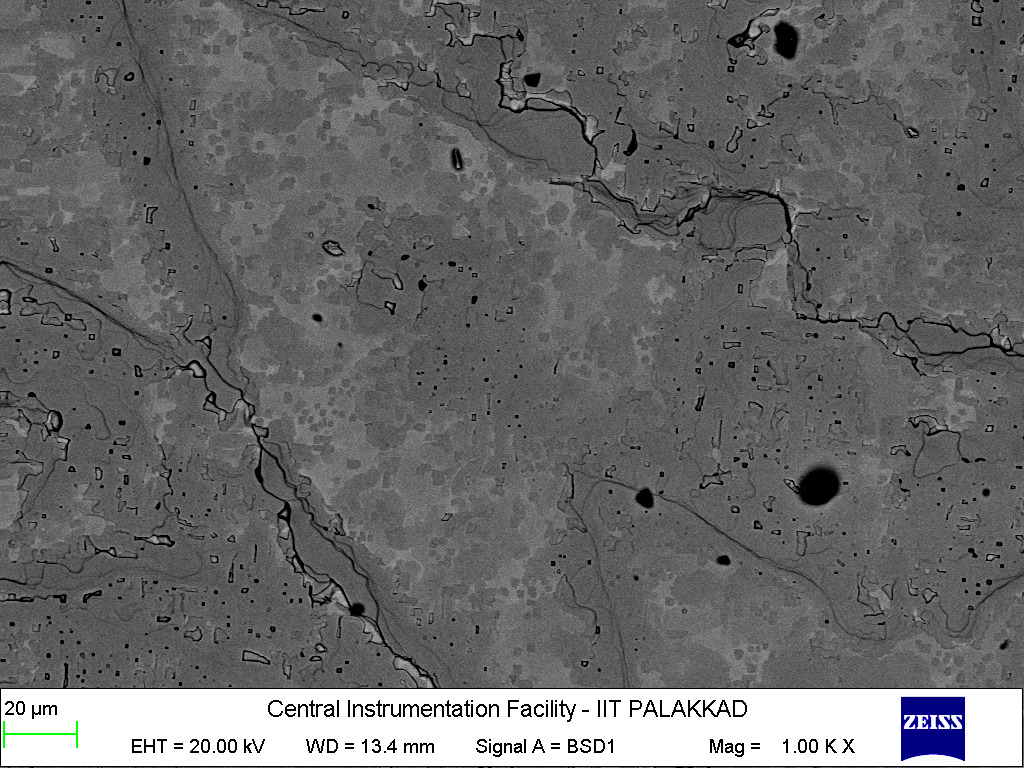}
     \put(3,68){\textcolor{white}{(b)}}
    \end{overpic}
\phantomcaption
\label{fig:3b}
  \end{subfigure}
  \hfill
  \begin{subfigure}[b]{0.33\textwidth}
    \begin{overpic}[width=\linewidth]{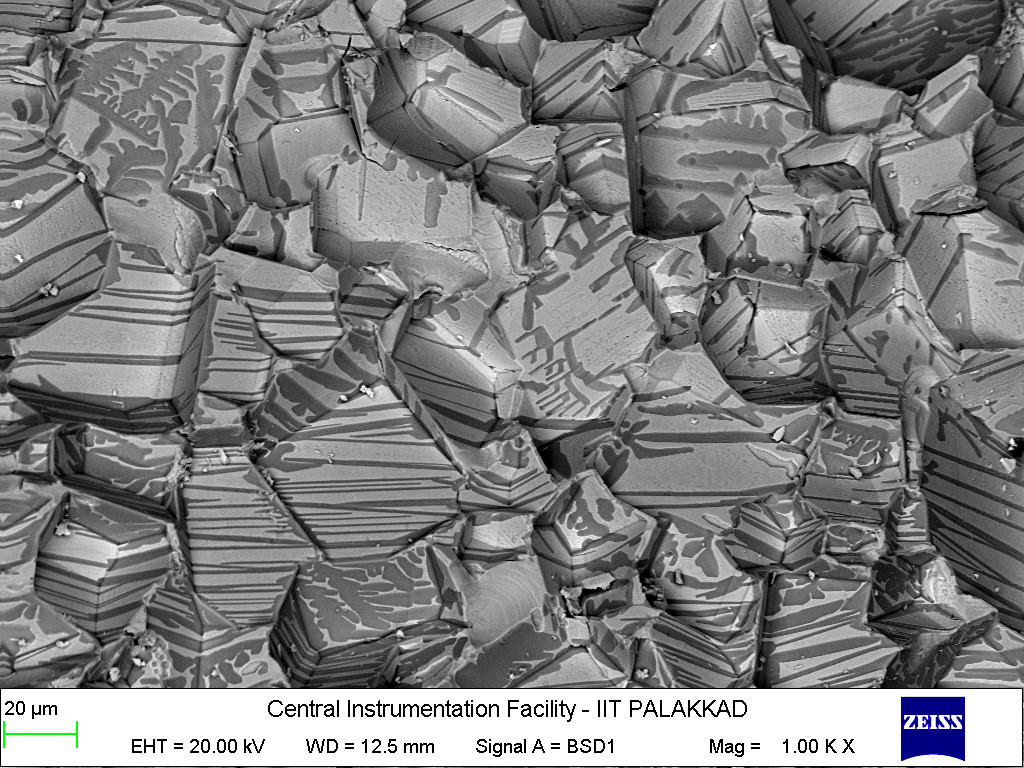}
     \put(3,68){\textcolor{white}{(c)}}
    \end{overpic}
\phantomcaption
\label{fig:3c}
  \end{subfigure}
  \caption{Back scattered electron images of (a) undoped, (b) 10$\%$ F-doped, and   (c) 20$\%$ F-doped BBO crystals }
\end{figure*}

\begin{table}[H]
    \centering
     \caption{Comparison between theoretical, undoped BBO, and $10\%$ fluorine-doped BBO lattice parameters.}
    \begin{tabular}{llll}
    \textbf{Parameter} & \textbf{Theoretical} & \textbf{BBO} & \textbf{F- doped BBO} \\
        \hline
        Space group & I 2/m & I 2/m & I 2/m \\
        $a$ (Å)     & 6.1908   & 6.1868(4)  & 6.1909(7)  \\
        $b$ (Å)     & 6.145    & 6.1349(5)  & 6.1347(7) \\
        $c$ (Å)     & 8.6785   & 8.6717(7)  & 8.6705(10) \\
        $V$ (Å$^3$) & 330.150  & 329.14(4)  & 329.29(3) \\
        $\beta$ (°) & 90.164   & 90.21(1)   & 90.207(21) \\
        $wR$ (\%)   & --       & 16.67      & 9.53 \\
    \end{tabular}
    
    \label{tab:1}
\end{table}

Crystals with flat surfaces were chosen for backscattered electron (BSE) imaging using a high-resolution FESEM. The images are shown in Figures [\ref{fig:3a}], [\ref{fig:3b}] and [\ref{fig:3c}] respectively, for undoped BBO, 10$\%$ F-doped, and 20$\%$ F-doped BBO.  In BSE imaging, contrast variations indicate differences in composition or phase; regions exhibiting different contrasts generally represent different phases within the sample \cite{manni2014effect}. The BSE images of undoped BBO and 10$\%$ F-doped BBO are seen to have a single contrast and indicate that the crystals are in single phase. In the BSE image of 20$\%$ F-doped BBO, alternative dark and light contrast lines are seen. It certainly indicates two phase samples. EDS measurement was carried out in the two phases. The lighter contrast regions represent a lower fluorine concentration and the darker contrast regions represent a higher fluorine concentration from EDS. The varying BSE imaging contrast, measured different composition on those different regions and the multiphase PXRD pattern of of 20\% F-doped BBO sample clearly indicates that it is a multiphase system.

\begin{figure*}[ht]
  \centering
  \begin{subfigure}[b]{0.33\textwidth}
    \begin{overpic}[width=\linewidth]{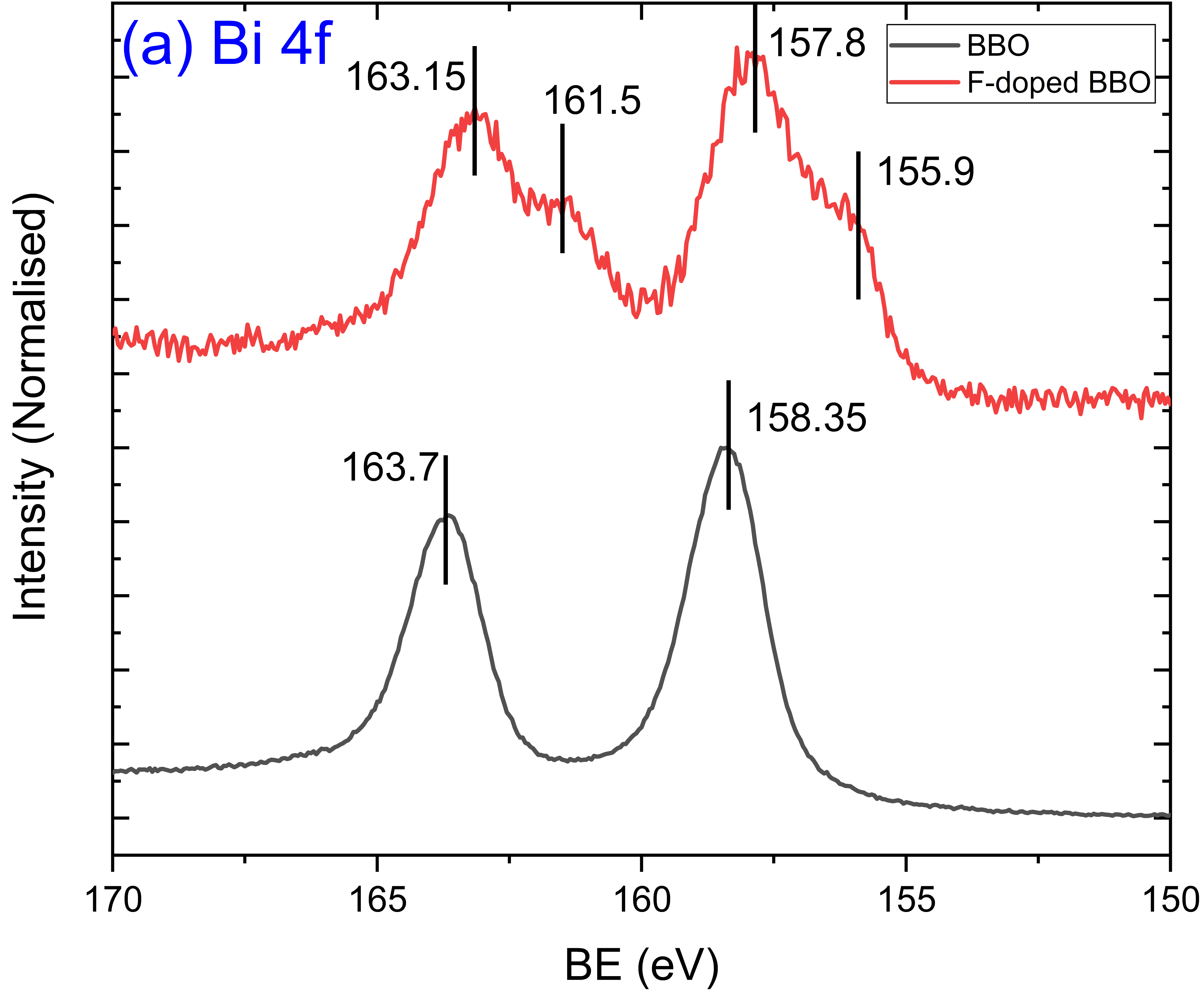}
    \end{overpic}
\phantomcaption
\label{fig:4a}
  \end{subfigure}
  \hfill
  \begin{subfigure}[b]{0.32\textwidth}
    \begin{overpic}[width=\linewidth]{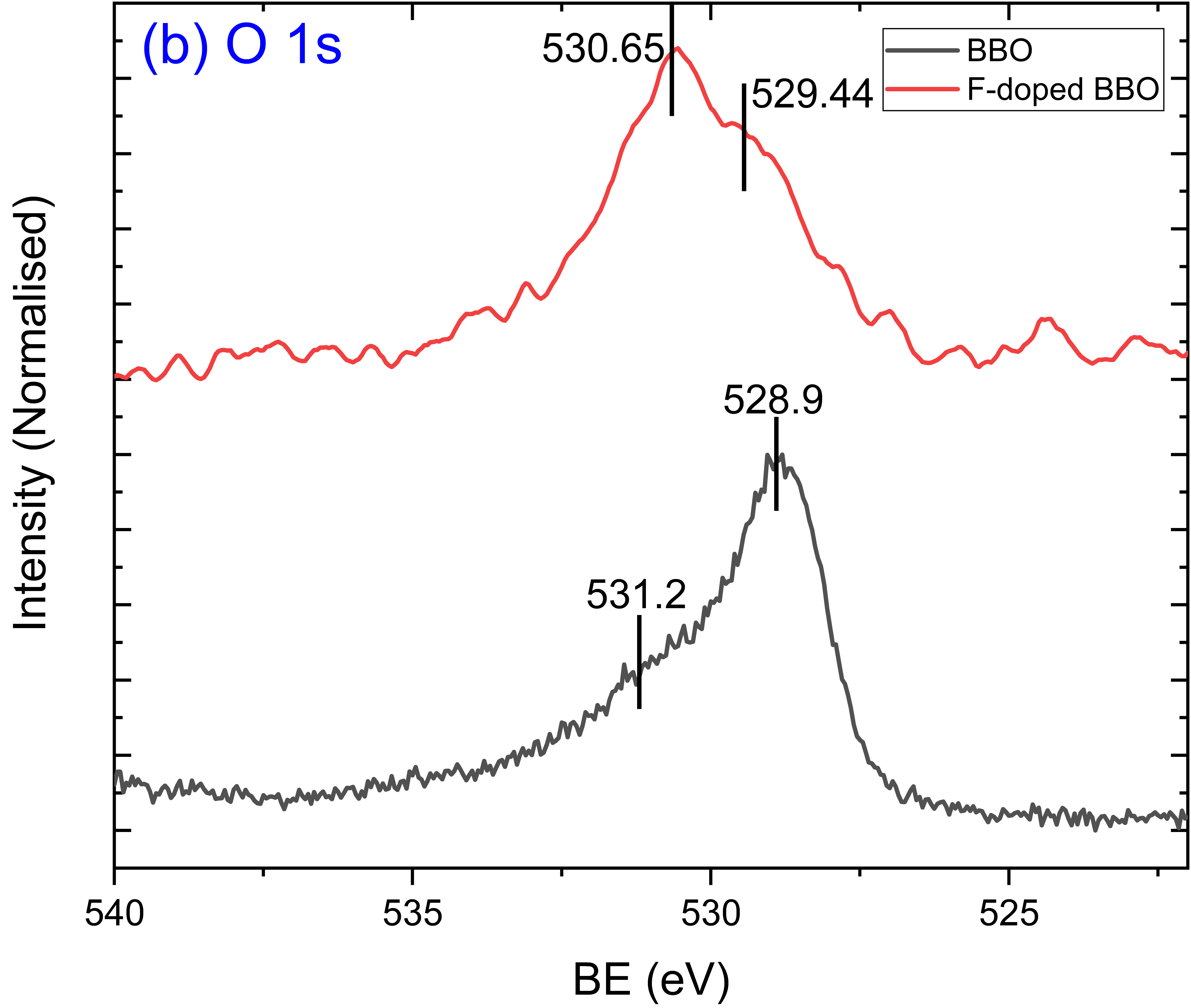}
    \end{overpic}
\phantomcaption
\label{fig:4b}
  \end{subfigure}
  \hfill
  \begin{subfigure}[b]{0.31\textwidth}
    \begin{overpic}[width=\linewidth]{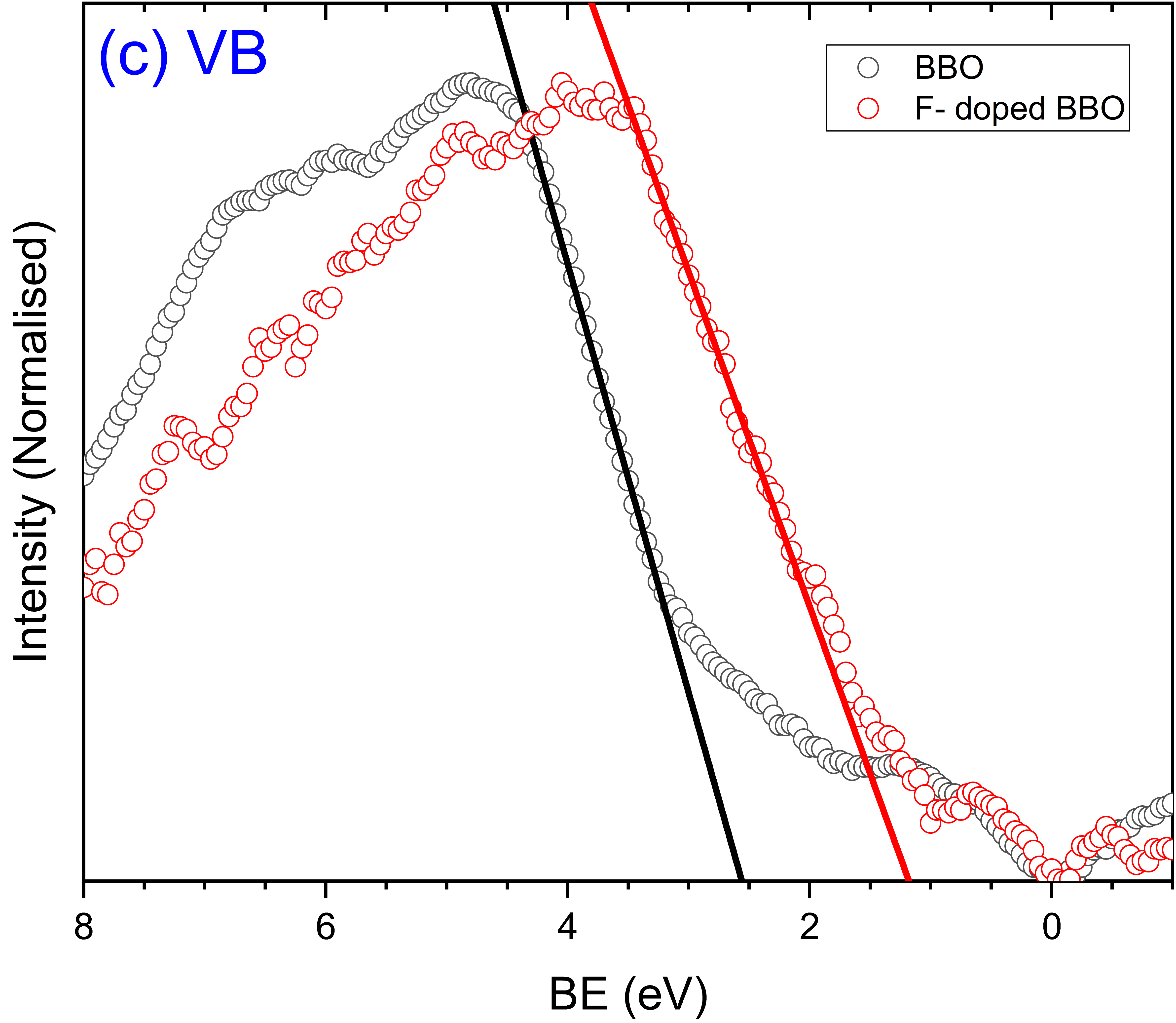}
    \end{overpic}
\phantomcaption
\label{fig:4c}
  \end{subfigure}
  \caption{X-ray photoelectron spectroscopy (XPS) of undoped  and 10$\%$ F-doped BBO. (a) Bi 4f core level spectra (b) O 1s core level spectra  (c) Valence band spectra with valence band maximum (VBM) extrapolated.}
\end{figure*}

Synchrotron X-ray photoelectron spectroscopy (XPS) studies on undoped and 10$\%$ F-doped BBO have been performed to study the effect of chemical doping on core and valence band spectra. The binding energy (BE) scale of the XPS core levels was calibrated with the BE position of the C 1s core level at 284.6 eV. The Bi 4f core level spectra of the undoped BBO in Figure [\ref{fig:4a}], shows spin-orbit split doublet peaks. In F-doped BBO, there is an extra peak in addition to the doublet seen in undoped BBO, as in Figure [\ref{fig:4a}]. This additional peak is associated with the Bi-F bonding environment and is reported by Vesto \textit{et al.}\cite{vesto2025growth} in their thin film studies of F-doped BBO on a sapphire substrate. In addition to this, the peak positions of the doped BBO are shifted to the lower binding energy, an indication of the electron doping occurring with the substitution of fluorine atoms at the oxygen-site. The O-1s spectra in Figure [\ref{fig:4b}] indicate contributions from both the lattice oxygen at 528.9 eV and the surrounding oxygen at 531.2 eV for BBO. In the F-doped BBO, the replacement of oxygen atoms with fluorine is evident as we find a decrease in the relative intensity of lattice oxygen compared to that of surrounding oxygen when compared with the undoped BBO.  Valence band (VB) spectra have been studied and the valence band maximum (VBM) has been estimated by fitting a solid line as shown in Figure [\ref{fig:4c}] and then extrapolated the fitted line to cut the binding energy scale. The shift of VBM toward a lower binding energy in the F-doped BBO compared with that in the undoped BBO indicates the electron doping in F-doped BBO.

\section{Summary and conclusions}
The single crystals of BBO and 10$\%$ F-doped BBO have been successfully grown by a single-step solid state technique and have been structurally characterized by XRD. The BSE images indicate that the crystals were grown in a single phase. Multiple phases are indicated in the 20$\%$ F-doped BBO BSE image. It can be concluded that the fluorine solubility level in
BBO could have been exceeded at the 20$\%$ fluorine doping
level using the solid-state technique, resulting in the formation
of multiple phases. The electron doping in F-doped BBO is confirmed from the XPS core and valence band spectra measurements.

In conclusion, we successfully synthesized pure and 10$\%$ F-doped BBO crystals using a new technique and confirmed electron doping. We have also determined the fluorine solubility level in single-crystalline bulk BaBiO$_3$.

\section*{Acknowledgements}
 The authors acknowledge the BL-10 beamline at Indus-2, RRCAT, Indore for the synchrotron XPS measurements and other experimental support from CIF, IIT Palakakd.  SM thanks Ms. Ruta Kulkarni and Crystal Growth Lab, DCMP\&MS, TIFR, Mumbai for the assistance in the Laue Diffraction experiment. The work was supported by the Department of Science and Technology (DST), India, through the INSPIRE Faculty Grant.

\appendix

\section{PXRD of 20\% F-doped BBO}
\begin{figure}[ht!]
     \centering
    \includegraphics[width=\linewidth]{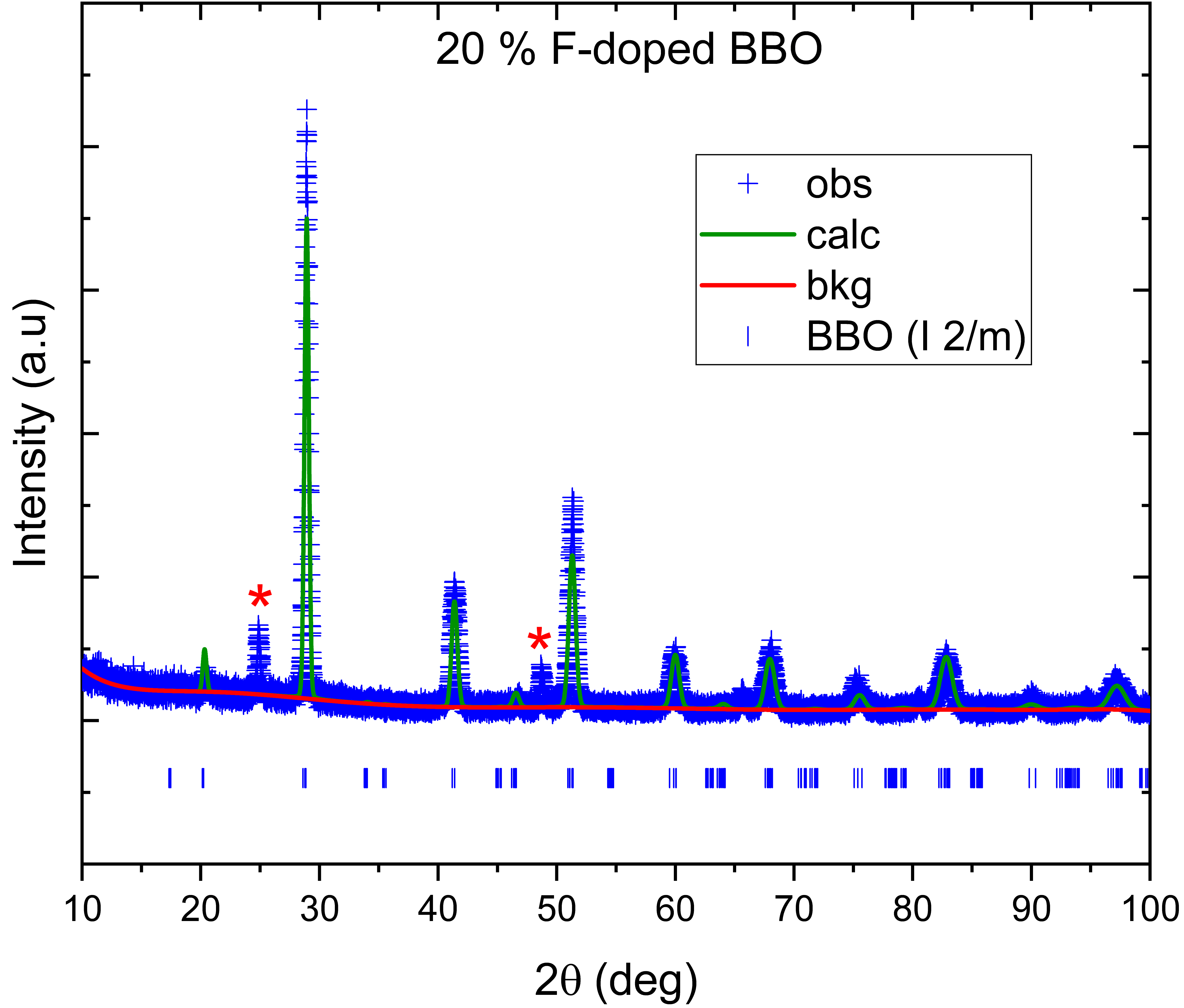}
    \caption{X-ray diffraction pattern of 20$\%$ F-doped BBO. The unmarked peaks correspond to the BBO phase, while the peak marked in red indicates the presence of barium fluoride as the impurity phase.}
    \label{fig:A5}
\end{figure}
The PXRD performed on 20$\%$ F-doped BBO indicates the presence of impurities. The structure is refined with monoclinic \textit{I2/m } crystal structure (SG No. 12) and the impurity peaks corresponding to barium fluoride (BaF$_2$) are marked with red stars in Figure [\ref{fig:A5}]. This confirms the multiphase nature of the 20$\%$ F-doped BBO sample.

\begin{table}[ht!]
\centering
\resizebox{\columnwidth}{!}{%
\begin{tabular}{|c|c|c|c|c|c|c|}
\hline
\textbf{Atom} & \textbf{Wyckoff} & \textbf{Phase} & \textbf{x/a} & \textbf{y/b} & \textbf{z/c} & \textbf{SOF} \\
\hline
Bi1 & 2a & BBO         & 0          & 0          & 0           & 0.932(10)  \\
    &    & F-doped BBO       & 0          & 0          & 0           & 0.84(4)    \\
\hline
Bi2 & 2c & BBO         & 0          & 0          & 0.5         & 1.012(15)  \\
    &    & F-doped BBO      & 0          & 0          & 0.5         & 0.94(4)    \\
\hline
Ba  & 4i & BBO         & 0.495(5)   & 0          & 0.2555(13)  & 1.009(8)   \\
    &    & F-doped BBO       & 0.5040(29) & 0          & 0.2554(21)  & 0.956(12)  \\
\hline
O1  & 4i & BBO         & 0.031(13)  & 0          & 0.272(21)   & 1          \\
    &    & F-doped BBO     & 0.045(12)  & 0          & 0.280(9)    & 0.936      \\
\hline
O2  & 8j & BBO         & 0.368(6)   & 0.166(8)   & -0.031(4)   & 1          \\
    &    & F-doped BBO    & 0.324(8)   & 0.176(10)  & -0.042(6)   & 0.936      \\
\hline
F1  & 4i & BBO           & -          & -          & -           & -          \\
    &    & F-doped BBO    & 0.045(12)  & 0          & 0.280(9)    & 0.064      \\
\hline
F2  & 8j & BBO        & -          & -          & -           & -          \\
    &    & F-doped BBO      & 0.324(8)   & 0.176(10)  & -0.042(6)   & 0.064      \\
\hline
\end{tabular}
}
\caption{Refined atomic positions and site occupancies for BBO and 10\% F-doped BBO.}
\label{AtomicCoordinates}
\end{table}

\newpage
\bibliographystyle{elsarticle-num}

\begin{thebibliography}{10}
\expandafter\ifx\csname url\endcsname\relax
  \def\url#1{\texttt{#1}}\fi
\expandafter\ifx\csname urlprefix\endcsname\relax\def\urlprefix{URL }\fi
\expandafter\ifx\csname href\endcsname\relax
  \def\href#1#2{#2} \def\path#1{#1}\fi

\bibitem{kane2005quantum}
C.~L. Kane, E.~J. Mele, Quantum spin hall effect in graphene, Physical review letters 95~(22) (2005) 226801.

\bibitem{konig2007quantum}
M.~Konig, S.~Wiedmann, C.~Brune, A.~Roth, H.~Buhmann, L.~W. Molenkamp, X.-L. Qi, S.-C. Zhang, Quantum spin hall insulator state in hgte quantum wells, Science 318~(5851) (2007) 766--770.

\bibitem{hasan2011three}
M.~Z. Hasan, J.~E. Moore, Three-dimensional topological insulators, Annu. Rev. Condens. Matter Phys. 2~(1) (2011) 55--78.

\bibitem{yan2013large}
B.~Yan, M.~Jansen, C.~Felser, A large-energy-gap oxide topological insulator based on the superconductor babio3, Nature Physics 9~(11) (2013) 709--711.

\bibitem{sleight2015bismuthates}
A.~W. Sleight, Bismuthates: Babio3 and related superconducting phases, Physica C: Superconductivity and its Applications 514 (2015) 152--165.

\bibitem{chouhan2018babio3}
A.~S. Chouhan, E.~Athresh, R.~Ranjan, S.~Raghavan, S.~Avasthi, Babio3: A potential absorber for all-oxide photovoltaics, Materials Letters 210 (2018) 218--222.

\bibitem{bouwmeester2021babio3}
R.~L. Bouwmeester, A.~Brinkman, Babio3—from single crystals towards oxide topological insulators, Reviews in Physics 6 (2021) 100056.

\bibitem{cava1988superconductivity}
R.~Cava, B.~Batlogg, J.~Krajewski, R.~Farrow, L.~J. Rupp, A.~White, K.~Short, W.~Peck, T.~Kometani, Superconductivity near 30 k without copper: the ba0. 6k0. 4bio3 perovskite, nature 332~(6167) (1988) 814--816.

\bibitem{sleight1993high}
A.~W. Sleight, J.~Gillson, P.~Bierstedt, High-temperature superconductivity in the bapb1- xbixo3 system, Solid State Communications 88~(11-12) (1993) 841--842.

\bibitem{khamari2019shifting}
B.~Khamari, B.~Nanda, Shifting of fermi level and realization of topological insulating phase in the oxyfluoride babio2f, Materials Research Express 6~(6) (2019) 066309.

\bibitem{vesto2025growth}
R.~Vesto, H.~Choi, K.~Kim, Growth of f-doped babio3 thin films via flow-limited field-injection electrostatic spraying toward experimental realization of wide bandgap topological insulator, Scripta Materialia 262 (2025) 116656.

\bibitem{shiozaki1993crystal}
I.~Shiozaki, H.~Ishii, Crystal growth and electronic structure of metal substituted babio3, Japanese Journal of Applied Physics 32~(S3) (1993) 686.

\bibitem{Toby:aj5212}
B.~H. Toby, R.~B. Von~Dreele, \href{https://doi.org/10.1107/S0021889813003531}{{{\it GSAS-II}: the genesis of a modern open-source all purpose crystallography software package}}, Journal of Applied Crystallography 46~(2) (2013) 544--549.
\newblock \href {https://doi.org/10.1107/S0021889813003531} {\path{doi:10.1107/S0021889813003531}}.
\newline\urlprefix\url{https://doi.org/10.1107/S0021889813003531}

\bibitem{pei1990structural}
S.~Pei, J.~Jorgensen, B.~Dabrowski, D.~Hinks, D.~Richards, A.~Mitchell, J.~Newsam, S.~Sinha, D.~Vaknin, A.~Jacobson, Structural phase diagram of the ba 1- x k x bio 3 system, Physical Review B 41~(7) (1990) 4126.

\bibitem{manni2014effect}
S.~Manni, S.~Choi, I.~Mazin, R.~Coldea, M.~Altmeyer, H.~O. Jeschke, R.~Valenti, P.~Gegenwart, Effect of isoelectronic doping on the honeycomb-lattice iridate a 2 iro 3, Physical Review B 89~(24) (2014) 245113.

\end{thebibliography}


\end{document}